\documentclass[letterpaper, 10 pt, conference]{ieeeconf}
\IEEEoverridecommandlockouts
\let\labelindent\relax 
\usepackage{enumitem}
\usepackage[T1]{fontenc}
\usepackage[utf8]{inputenc}
\usepackage{graphicx}
\usepackage{amsmath,mathtools,booktabs}
\usepackage{amssymb}
\usepackage[dvipsnames]{xcolor} 
\usepackage{url}
\usepackage[ruled,linesnumbered]{algorithm2e}   
\usepackage{tikz}                               
\usepackage{pgfplots}                           
\pgfplotsset{compat=1.17}
\usepackage{scalefnt}                           
\usepackage{subcaption}
\usepackage{commath}

\usepackage[font=footnotesize]{caption}

\definecolor{clrAMPIP}{rgb}{1,0.5,0.1}%
\definecolor{clrTDCR}{rgb}{0.6,0.8,0.2}%
\definecolor{clrOAADMM}{rgb}{0,0.5,0.8}%

\usepackage{amsthm}

\newtheorem{theorem}{Theorem}

\newtheorem{lemma}{Lemma}

\graphicspath{{./image/}}
\usepackage{cleveref}
\usepackage{bm}
\newcommand{\vm}[1]{\boldsymbol{#1}} 
\newcommand{\mc}[1]{\mathcal{#1}} 
\newcommand{\mb}[1]{\mathbb{#1}} 
\newcommand{\ouset}[3]{\underset{#2}{\overset{#1}{#3}}} 
\newcommand{\txtn}[1]{\textnormal{#1}}

\newcommand{\invm}{\scalebox{0.75}[1.0]{$-$}} 
\usetikzlibrary{arrows.meta,positioning,patterns,shapes,calc}
\tikzset{   
	>={Latex[width=2mm,length=2mm]},
	block/.style = {draw, fill=white, rectangle, minimum height=3em, minimum width=6em},
	mux/.style = {draw, fill=black, rectangle, text centered, minimum height = 1cm, minimum width = 1mm,inner sep=0pt},
	dot/.style = {draw, fill=black, circle,minimum width=1mm, minimum height=1mm,inner sep=0pt},
}
\title{\LARGE \bf Flexible MPC-based Conflict Resolution Using Online Adaptive ADMM}

\author{Jerry An$^{1,2}$, Giulia Giordano$^3$ and Changliu Liu$^1$
\thanks{$^1$ Robotics Institute, Carnegie Mellon University, Pittsburgh, USA, {\tt\small jerryan.x@gmail.com, cliu6@andrew.cmu.edu}.}%
\thanks{$^2$ Delft Center for Systems and Control, Delft University of Technology, Delft, the Netherlands, {\tt\small jerryan.x@gmail.com}.}%
\thanks{$^3$ Department of Industrial Engineering, University of Trento, Italy, {\tt\small giulia.giordano@unitn.it}.}
}

\begin{document}

\maketitle
\begin{abstract}
Decentralized conflict resolution for autonomous vehicles is needed in many places where a centralized method is not feasible, e.g., parking lots, rural roads, merge lanes, etc. However, existing methods generally do not fully utilize optimization in decentralized conflict resolution. We propose a decentralized conflict resolution method for autonomous vehicles based on a novel extension to the Alternating Directions Method of Multipliers (ADMM), called Online Adaptive ADMM (OA-ADMM), and on Model Predictive Control (MPC). OA-ADMM is tailored to online systems, where fast and adaptive real-time optimization is crucial, and allows the use of safety information about the physical system to improve safety in real-time control. 
We prove convergence in the static case and give requirements for online convergence. Combining OA-ADMM and MPC allows for robust decentralized motion planning and control that seamlessly integrates decentralized conflict resolution. The effectiveness of our proposed method is shown through simulations in CARLA, an open-source vehicle simulator, resulting in a reduction of 47.93\% in mean added delay compared with the next best method.
\end{abstract}

\section{Introduction} 
When designing fully autonomous vehicles, reducing traffic congestion is a crucial goal. Intersections are a major contributor to traffic delays and accidents, hence autonomous vehicles need to be equipped to deal with them efficiently \cite{rahmati_towards_2017}. 
Since intersections often lack the infrastructure required to centrally resolve the conflicts \cite{khan_smart_2019}, autonomous vehicles must be able to resolve conflicts without any external infrastructure. Navigating unmanaged intersections using decentralized policies is challenging due to the risk of deadlocks or accidents: communication and conflict resolution protocols among autonomous vehicles are needed.

Traditional approaches for conflict resolution include heuristic intersection protocols using various priority policies \cite{azimi_intersection_2012,azimi_reliable_2013}, which  allow real-time adjustments of the priorities and can utilize constrained optimal control to improve their performance \cite{liu_distributed_2018}.
Online distributed motion planning for the formation control of multi-agent systems, based on a single-iteration receding horizon approach, is proposed in \cite{van_parys_online_2016}, while \cite{van_parys_distributed_2017} also considers inter-vehicle collision avoidance through the use of separating hyperplanes. The ADMM (Alternating Directions Method of Multipliers) variant, introduced in \cite{zheng_fast_2017} for autonomous vessels, utilizes a central coordinator and claims to improve the convergence rate by iteratively adding approximated collision avoidance constraint. A method similar to that of \cite{van_parys_online_2016}, \cite{van_parys_distributed_2017} is introduced in \cite{rey_fully_2018}, utilizing a linearized collision avoidance constraint and incorporating deadlock-protection. The nonlinear MPC-based approach in \cite{firoozi_distributed_2020} reduces the need for linearized constraints. The distributed MPC approach in \cite{chen_distributed_2018} incorporates the residual balancing method from \cite{he_alternating_2000}, while \cite{9007450} utilizes distributed trajectory optimization based on a decomposition technique that avoids communication between agents until convergence. 

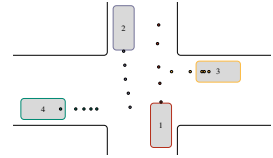
\begin{figure}[t]
	\centering
	\resizebox{0.4\linewidth}{!}{
		\begin{tikzpicture}[auto] 
		\clip (-5.5,-2.75) rectangle + (11,6.5); 
		\node (rect) at (5,5) [draw,thick,minimum width = 7cm, minimum height = 7cm, rounded corners] {};
		\node (rect) at (5,-5) [draw,thick,minimum width = 7cm, minimum height = 7cm, rounded corners] {};
		\node (rect) at (-5,5) [draw,thick,minimum width = 7cm, minimum height = 7cm, rounded corners] {};
		\node (rect) at (-5,-5) [draw,thick,minimum width = 7cm, minimum height = 7cm, rounded corners] {};
		\node (rect) at (0.8,-1.5) [name = agent1, fill = gray!30, draw= BrickRed,very thick,minimum width = 0.9cm, minimum height = 1.9cm, rounded corners] {1};	
		\node (rect) at (-0.8,2.675) [name = agent2, fill = gray!30, draw=CadetBlue,very thick,minimum width = 0.9cm, minimum height = 1.9cm, rounded corners] {2};	
		\node (rect) at (3.25,0.8) [name = agent3, fill = gray!30, draw=Dandelion,very thick,minimum width = 1.9cm, minimum height = 0.9cm, rounded corners] {3};				
		\node (rect) at (-4.25,-0.8) [name = agent4, fill = gray!30, draw=PineGreen,very thick,minimum width = 1.9cm, minimum height = 0.9cm, rounded corners] {4};			
		\node [draw, fill=BrickRed, circle,minimum width=1mm, minimum height=1mm,inner sep=0pt] at ([shift=({0cm,1cm})]agent1)  {};
		\node [draw, fill=BrickRed, circle,minimum width=1mm, minimum height=1mm,inner sep=0pt] at ([shift=({-0.05cm,1.8cm})]agent1)  {};
		\node [draw, fill=BrickRed, circle,minimum width=1mm, minimum height=1mm,inner sep=0pt] at ([shift=({-0.1cm,2.5cm})]agent1)  {};
		\node [draw, fill=BrickRed, circle,minimum width=1mm, minimum height=1mm,inner sep=0pt] at ([shift=({-0.15cm,3.5cm})]agent1)  {};
		\node [draw, fill=BrickRed, circle,minimum width=1mm, minimum height=1mm,inner sep=0pt] at ([shift=({-0.1cm,4.3cm})]agent1)  {};
		\node [draw, fill=CadetBlue, circle,minimum width=1mm, minimum height=1mm,inner sep=0pt] at ([shift=({0cm,-1cm})]agent2)  {};
		\node [draw, fill=CadetBlue, circle,minimum width=1mm, minimum height=1mm,inner sep=0pt] at ([shift=({0.05cm,-1.6cm})]agent2)  {};
		\node [draw, fill=CadetBlue, circle,minimum width=1mm, minimum height=1mm,inner sep=0pt] at ([shift=({0.075cm,-2.2cm})]agent2)  {};
		\node [draw, fill=CadetBlue, circle,minimum width=1mm, minimum height=1mm,inner sep=0pt] at ([shift=({0.2cm,-2.8cm})]agent2)  {};
		\node [draw, fill=CadetBlue, circle,minimum width=1mm, minimum height=1mm,inner sep=0pt] at ([shift=({0.275cm,-3.4cm})]agent2)  {};				
		\node [draw, fill=Dandelion, circle,minimum width=1mm, minimum height=1mm,inner sep=0pt] at ([shift=({-0.4cm,0cm})]agent3)  {};
		\node [draw, fill=Dandelion, circle,minimum width=1mm, minimum height=1mm,inner sep=0pt] at ([shift=({-0.6cm,0cm})]agent3)  {};
		\node [draw, fill=Dandelion, circle,minimum width=1mm, minimum height=1mm,inner sep=0pt] at ([shift=({-0.7cm,0cm})]agent3)  {};
		\node [draw, fill=Dandelion, circle,minimum width=1mm, minimum height=1mm,inner sep=0pt] at ([shift=({-1.2cm,0cm})]agent3)  {};
		\node [draw, fill=Dandelion, circle,minimum width=1mm, minimum height=1mm,inner sep=0pt] at ([shift=({-2.0cm,0cm})]agent3)  {};
		\node [draw, fill=PineGreen, circle,minimum width=1mm, minimum height=1mm,inner sep=0pt] at ([shift=({0.75cm,0cm})]agent4)  {};
		\node [draw, fill=PineGreen, circle,minimum width=1mm, minimum height=1mm,inner sep=0pt] at ([shift=({1.4cm,0cm})]agent4)  {};
		\node [draw, fill=PineGreen, circle,minimum width=1mm, minimum height=1mm,inner sep=0pt] at ([shift=({1.75cm,0cm})]agent4)  {};
		\node [draw, fill=PineGreen, circle,minimum width=1mm, minimum height=1mm,inner sep=0pt] at ([shift=({2.05cm,0cm})]agent4)  {};
		\node [draw, fill=PineGreen, circle,minimum width=1mm, minimum height=1mm,inner sep=0pt] at ([shift=({2.3cm,0cm})]agent4)  {};
		
		\end{tikzpicture}%
	}
	\caption{Example of conflict resolution using OA-ADMM and MPC. The colored dots represent the MPC trajectories of each vehicle.}%
	\label{fig:oa-admm_FigureExample}%
	\vspace{-20pt}
\end{figure}


The decentralized protocols in \cite{azimi_intersection_2012,azimi_reliable_2013,liu_distributed_2018} require prior knowledge of the environment and cannot adjust the vehicles' trajectories. The online distributed motion planning techniques proposed in \cite{van_parys_online_2016,van_parys_distributed_2017,rey_fully_2018,firoozi_distributed_2020} do not allow deadlock resolution or adaptive penalty parameters, making safety during real-time implementation questionable. The distributed MPC approach from \cite{chen_distributed_2018} does not utilize the adaptive penalty parameter to improve the system safety. The distributed trajectory optimization technique from \cite{9007450} requires performing the optimization steps until convergence and the decomposition method has no convergence proofs, making the method unsuitable for autonomous vehicles.

We propose a novel MPC-based method for decentralized conflict resolution that relies on our proposed Online Adaptive ADMM (OA-ADMM) algorithm, which improves efficiency because it allows trajectory deviation. When ADMM cannot be performed until convergence due to the required control frequency, feasibility results can be poor to the extent that safety is an issue, in addition convergence can also be poor when results from the previous control step are not utilized. OA-ADMM solves this using two user-designed functions: the similarity function, which is a forgetting factor between two time steps of the online system, and the adaptation function, which adjusts the penalty parameters between updates. In our application we design the similarity function and adaptation function based on the physical safety of the system, increasing both values when distance between planned trajectories decrease.
Our main contributions are:
\begin{itemize}[noitemsep,topsep=0pt,parsep=0pt,partopsep=0pt]
    \item Unification of online application of ADMM under one framework (OA-ADMM) 
    \item Proposal of a physical safety based adaptation function to improve online robustness and safety.
    \item Application of OA-ADMM and MPC, achieving trajectory deviation in decentralized conflict resolution, with fewer requirement in terms of prior knowledge.
\end{itemize}


\section{Problem Formulation} \label{sec:problemformulation}

When approaching conflict resolution as a centralized optimization problem, it takes the form:
\begin{equation}
    \begin{aligned}
    \underset{\vm{x}}{\min} & \quad \ouset{M}{i=1}{\sum}J_i(\vm{x})                               \\
    \text{s.t.}                    & \quad \vm{x}_i \in \mc{X}_i^f,                                     \\
    & \quad d(\vm{x}_i,\vm{x}_j) \geq 0, \forall j \in \mc{N}_i, \quad \forall i \in \{1,...,N\},
    \end{aligned}
\label{eq:_systemObjective}
\end{equation}
where $\vm{x}_i$ is the state vector for agent $i$ (e.g. for a kinematic bicycle model $\vm{x}_i=[\txtn{x},\txtn{y},v,a,\beta]^\top$, with \txtn{x,y} the coordinates in the local coordinate frame, $a$ the input acceleration, and $\beta$ the steering angle); $\mc{X}_i^f$ is the feasible set for $\vm{x}_i$ including all trajectories that adhere to the system dynamics, input constraint, and environmental collision avoidance constraints; $d(\vm{x}_i,\vm{x}_j)$ is the distance function between two capsules, 
and $\mc{N}_i$ is the set of vehicles neighboring $i$. A capsule is a line segment of length $l$ inflated with a radius $r$.

The optimization problem \eqref{eq:_systemObjective} is difficult to solve, being nonlinear and nonconvex, and the constraints couple the states of more agents. For real time applications, a fast (<100ms) solution is required, making a centralized approach unpractical as the problem does not scale well.
Due to the coupling in the constraints and objective function, primal decomposition cannot be applied to problem \eqref{eq:_systemObjective}, which is shown to be NP-hard in \cite{colombo_efficient_2012}.
For real-time optimization-based conflict resolution, problem \eqref{eq:_systemObjective} can be reformulated into the MPC-based finite horizon form, with state vector $\vm{x}\in\mb{R}^{N_x}$, where $N_x$ is the length of the finite horizon time the amount of states per time step. Also, to be able to apply our proposed Online Adaptive ADMM approach to problem \eqref{eq:_systemObjective}, we need to reformulate it as a general ADMM problem: 
\begin{equation} \label{eq:oa-admm_meth_MainProblem}
	\underset{\vm{x},\vm{z}}{\min} \quad f(\vm{x}) + g(\vm{z})     \quad
	\txtn{s.t.}       \quad \vm{Ax}+\vm{Bz}=\vm{c},
\end{equation}
where $f(\vm{x})$ and $g(\vm{z})$ are convex functions, $\vm{x}\in\mb{R}^n,\vm{z}\in\mb{R}^m,\vm{A}\in\mb{R}^{p\times n},\vm{B}\in\mb{R}^{p\times m}$, and $\vm{c}\in\mb{R}^{p}$. 


\section{Online Adaptive ADMM (OA-ADMM)} \label{sec:OA-ADMM} 
\begin{figure}[t]
	\centering
	\resizebox{0.90\linewidth}{!}{
	\begin{tikzpicture}[auto, node distance=4.5cm] 
	\node [block, name=x-update,align=center,minimum height = 1.2cm] {$x$-update:\\ \Cref{eq:oa-admm_meth_X-Update}};
	\node [block, right= 1cm of x-update, name=z-update,align=center,minimum height = 1.2cm] {$z$-update:\\ \Cref{eq:oa-admm_meth_Z-Update}};
	\node [block, right= 0.8cm of z-update, name=lambda-update,align=center,minimum height = 1.2cm] {$\lambda$-update:\\ \Cref{eq:oa-admm_meth_Lambda-Update}};
	\node [block, right=0.8cm of lambda-update, name= rho-update, align=center,minimum height = 1.2cm] {$\rho$-update:\\ \Cref{eq:oa-admm_meth_Rho-Update}};
	\node [mux, above=0.7cm of lambda-update, xshift= -2.15cm, name=mux] {};

	\draw [dashed,->] (lambda-update) -- node[] {$\vm{\lambda}^{k+1}$} (rho-update);
	\draw [dashed,->] (x-update.east) -| node[] {} ++(0.45cm,-1.1cm) -| ([xshift= -0.35cm] rho-update.south);	
	\draw [dashed,->] (z-update.east) -| node[] {} ++(0.45cm,-1.6cm) -| ([xshift= 0.35cm] rho-update.south);	
	
	\draw [->] (x-update) -- node[] {$\vm{x}^{k+1}$} (z-update);
	\draw [->] (x-update.east) -| node[] {} ++(0.45cm,-1.1cm) -| (lambda-update);			
	\draw [->] (z-update) -- node[] {$\vm{z}^{k+1}$} (lambda-update);
	\draw [->] ([xshift=-0.35cm]lambda-update.north) -- node[left] {$\vm{\lambda}^{k+1}$} ++(0,0.5cm) |- ([yshift=-0.25cm] mux.east);
	\draw [->] (rho-update.north) -- node[right] {$\vm{\rho}^{k+1}$} ++(0,1cm) |- ([yshift=0.25cm] mux.east);

	\draw [->] (rho-update.north) -- ++(0,1.46cm) -| ([xshift=0.35cm] lambda-update.north);	
	\draw [->] (mux.west) -| ( z-update);
	\draw [->] (mux.west) -- node[above] {$\vm{\lambda}^{k+1},\vm{\rho}^{k+1}$} ++(-1.9cm,0) -| ( x-update);	
	
	\node [dot,right= 0.395cm of x-update] {};
	\node [dot,right= 0.395cm of z-update] {};
	\node [dot,above= 1.15cm of z-update] {};
	\node [dot,yshift = -0.4925cm] at (lambda-update.south) {};
	\node [dot,above= 1.4cm of lambda-update,xshift=0.35cm] {};
	\end{tikzpicture}%
	}
	\caption{Diagram of OA-ADMM steps, dashed arrows resemble potential, but not necessary, inputs.}%
	\label{fig:oa-admm_meth_Diagram}%
	\vspace{-20pt}
\end{figure}
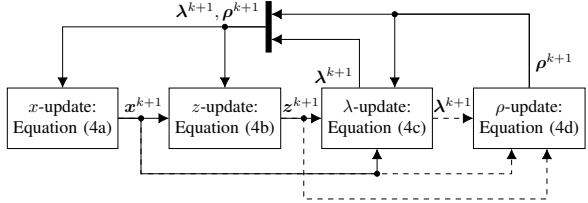
We propose a novel ADMM-based method, OA-ADMM, tailored to problems that require the ability to adapt constraint feasibility in real-time and for which conventional optimization methods cannot achieve convergence at the desired control frequency.
To be suitable for online optimization, OA-ADMM applies an ADMM-based strategy that yields admissible online results. Also, an adaptive penalty parameter $\rho$ is designed to allow prioritization of the constraint violations in online results.
As we will see, OA-ADMM guarantees improved robustness through the user-designed adaptation function $\phi$ and similarity function $\mu$, along with the vectorization of the penalty parameter $\rho$.

Similar to ADMM, the coupled constraints are integrated into an augmented Lagrangian to separate the problem, i.e.
\begin{equation} \label{eq:oa-admm_meth_MainLagrangian}
\begin{aligned}
	\mc{L}_{\vm{\rho}}(\vm{x},\vm{z},\vm{\lambda})  = f(\vm{x}) +g(\vm{z})  + \vm{\lambda}^\top (\vm{Ax}+\vm{Bz}-\vm{c})\\+\  \frac{1}{2}\Vert \vm{R}(\vm{Ax}+\vm{Bz}-\vm{c})\Vert_2^2.
\end{aligned}
\end{equation}
where $\vm{R}\in\mb{R}^{p\times p}$ is a diagonal matrix with $\txtn{diag}(\vm{R})=\vm{\rho}^{\circ\frac{1}{2}}$. The operator denoted by $(\cdot)^\circ$ is the \textit{Hadamard power} (or element-wise power). 
Using a penalty vector $\vm{\rho}$ we can adjust the penalty for each element of the primal residual $\vm{r}$, where the primal residual is $\vm{r}^{k+1}=\vm{Ax}^{k+1}+\vm{Bz}^{k+1}-\vm{c}$. The dual residual is $\vm{s}^{k+1}=\vm{A}^\top \vm{\rho}^k \circ \vm{B}(\vm{z}^{k+1}-\vm{z}^{k})$.

OA-ADMM requires two optimization steps, a Lagrangian multiplier update, and a $\rho$ update:
\begin{subequations} \label{eq:oa-admm_meth_Steps}
	\begin{equation} \label{eq:oa-admm_meth_X-Update}
	\vm{x}^{k+1}:= \txtn{arg}\ \underset{\vm{x}}{\txtn{min}} \  \mc{L}_{\vm{\rho}}(\vm{x},\vm{z}^{k},\vm{\lambda}^{k},\vm{\rho}^k),
	\end{equation}
	\begin{equation} \label{eq:oa-admm_meth_Z-Update}
	\vm{z}^{k+1}:= \txtn{arg}\ \underset{\vm{z}}{\txtn{min}} \  \mc{L}_{\vm{\rho}}(\vm{x}^{k+1},\vm{z},\vm{\lambda}^{k},\vm{\rho}^k),
	\end{equation}
	\begin{equation} \label{eq:oa-admm_meth_Lambda-Update}
	\vm{\lambda}^{k+1}:= \mu(\cdot) \vm{\lambda}^k + \vm{\rho}^{k} \circ \vm{r}^{k+1},
	\end{equation}
	\begin{equation} \label{eq:oa-admm_meth_Rho-Update}
	\vm{\rho}^{k+1}:= \phi(\cdot),
	\end{equation}
\end{subequations}
where $\phi$ is the adaptation function (see \Cref{ssec:oa-admm_meth_DesignPhi}) and $\mu$ is the similarity function (see \Cref{ssec:croa_meth_DesignMu}). The structure of OA-ADMM is visualized in \Cref{fig:oa-admm_meth_Diagram}.

We now provide convergence results and proofs. In the static case, with the problem assumed to be time-independent, using a similar approach to \cite[Appendix A]{boyd_distributed_2011} we prove that the OA-ADMM scheme converges as $k\rightarrow \infty$.

To prove convergence, we assume that the original Lagrangian has a saddle point at ($\vm{x}^\star,\vm{z}^\star,\vm{\lambda}^\star$) and we propose a candidate Lyapunov function $V$, such that $V\geq 0$ and $V=0$ only at the saddle point:
\begin{equation} \label{eq:oa-admm_cprf_static_Lyapunov}
V^k = \Vert\vm{R}^{\circ \invm 1}(\vm{\lambda}^{k}-\vm{\lambda}^{\star}) \Vert_2^2 + \Vert \vm{R} \vm{B}(\vm{z}^k-\vm{z}^\star)\Vert_2^2,
\end{equation}

We also state some preliminary lemmas. 
\begin{lemma}[Converging penalty parameter] \label{lm:oa-admm_cprf_static_Lem4}
For ADMM to convergence with an adaptive penalty vector $\vm{\rho}$, it is necessary that $\vm{\rho}^k \rightarrow \vm{\rho}^\star$ as $k\rightarrow \infty$, and that all elements of $\vm{\rho}$ are positive.
\end{lemma}
 Convergence for an adaptive penalty parameter is proven in \cite{he_alternating_2000} when $\rho$ converges to a certain $\rho^\star$. The requirement for all elements of $\vm{\rho}$ to be positive is given in \cite[Section 3.4.2]{boyd_distributed_2011}.
The following lemmas have been proven for ADMM with a conventional penalty parameter in \cite[Appendix A]{boyd_distributed_2011} and can be easily extended to the case when the parameter is replaced by a vector (which leads to small changes and the need of using the Hadamard operator).
\begin{lemma}[Objective suboptimality bounds] \label{lm:oa-admm_cprf_static_Lem}
The suboptimality of the objective function $p$ at step $k+1$, i.e. the difference between $p^{k+1}$ and the saddle point $p^\star$, is bounded as:
$-{\vm{\lambda}^\star}^\top \vm{r}^{k+1} \leq p^{k+1} - p^{\star} \leq -{\vm{\lambda}^{k+1}}^\top \vm{r}^{k+1} - \left( \vm{\rho}^k \circ \vm{B}(\vm{z}^{k+1}-\vm{z}^{k}) \right)^\top \left( -\vm{r}^{k+1} + \vm{B}(\vm{z}^{k+1}-\vm{z}^\star )\right)$
\end{lemma}
\begin{lemma}[Lyapunov decrease] \label{lm:oa-admm_cprf_static_Lem1}
The Lyapunov function $V$ in \eqref{eq:oa-admm_cprf_static_Lyapunov} decreases at each iteration as:
\begin{equation} \label{eq:oa-admm_cprf_static_VecIneq1}
V^{k+1} \leq V^k - \Vert \vm{R}^k \vm{r}^{k+1} \Vert_2^2 - \Vert \vm{R}^k \vm{B}(\vm{z}^{k+1}-\vm{z}^k)  \Vert_2^2.
\end{equation}
\end{lemma}

The convergence results for the static case can be summarized in the following theorem.
\begin{theorem} \label{th:oa-admm_cprf_stat_Theorem}
When applying the OA-ADMM algorithm in \eqref{eq:oa-admm_meth_Steps}, given closed, proper, and convex (See  \cite[Section 3.2]{boyd_distributed_2011}) functions $f$ and $g$, a saddle point $p^\star$ in the Lagrangian $\mc{L}$, and a converging $\vm{\rho}^k$, the primal and dual residuals converge to zero, $\vm{r}^k \rightarrow 0$ and $\vm{s}^k \rightarrow 0$, and the objective function value converges to its saddle point, $p^k \rightarrow p^\star$.
\end{theorem}
For proof, see \Cref{apx:proof_theorem1}.
To prove online convergence, we analyze the change of the system compared with the convergence of OA-ADMM. We distinguish between the OA-ADMM iteration parameter $k$ and the real time time step $t$: $\vm{x}^\star(t)$ is $\vm{x}$ at the saddle point for time $t$, whereas $\vm{x}^{k}(t)$ is $\vm{x}$ at iteration $k$ at time step $t$. 

Whilst the static convergence given in \Cref{th:oa-admm_cprf_stat_Theorem} proves residual convergence and objective convergence as $k\rightarrow \infty$, this result does not directly extend to the online case: for online systems, $\vm{x}^\star(t) = \vm{x}^\star(t+\delta t)$ cannot be assumed unless $\delta \rightarrow 0$, i.e. the control time step is very small. The change in optimum may be larger than the rate of convergence of OA-ADMM, which would result in each iteration converging towards its optimum, whilst never reaching the optimum at the next time step.
We therefore define online convergence to be that OA-ADMM can always converge towards the online optimum as $k,t\rightarrow\infty$.
This convergence requirement can be written as $\vm{x}^\star(t)=  \vm{x}^{k+1}(t),$ and $\vm{z}^\star(t)=  \vm{z}^{k+1}(t), \textnormal{ for } k,t \rightarrow \infty$.
Since proving convergence depends on the rate of change of the optimum over time, the convergence rate of OA-ADMM needs to always dominate the change of optimum as $k,t\rightarrow \infty$, i.e. $
	\Vert \vm{x}^{k_N}(t) - \vm{x}^\star(t)\Vert_2 - \Vert \vm{x}^{k_0}(t) - \vm{x}^\star(t)\Vert_2 >  \Vert \vm{x}^{\star}(t+\delta t) - \vm{x}^\star(t)\Vert_2$,
$\Vert \vm{z}^{k_N}(t) - \vm{z}^\star(t)\Vert_2 - \Vert \vm{z}^{k_0}(t) - \vm{z}^\star(t)\Vert_2 > \Vert \vm{z}^{\star}(t+\delta t) - \vm{z}^\star(t)\Vert_2$, $\forall t$,
where $k_0$ (resp. $k_N$) is the first (resp. last) iteration per time step. 
However, we are not aware of methods that can guarantee this in general for ADMM based methods; online convergence analysis should be performed on a system specific basis. In practice, to improve online convergence, it is advised to increase the amount of OA-ADMM iterations per control step, or the control frequency.
\section{OA-ADMM MPC} \label{sec:OA-ADMM_MPC} 
OA-ADMM can be applied to the MPC problem formulation \eqref{eq:_systemObjective} provided that a finite horizon is used. Similarly to \cite{van_parys_online_2016}, this is done by introducing a slack variable $\vm{z}$ and an equality constraint $\vm{x}=\vm{z}$:
\begin{equation} \label{eq:oacr_meth_slack}
\begin{aligned}
\underset{\vm{x},\vm{z_{ij}}}{\min} & \quad \ouset{N}{i=1}{\sum}J_i(x)                                          \\
s.t.                                & \quad \vm{x}_i \in \mc{X}_i^f,                                                \\
& \quad d(\vm{z}_i,\vm{z}_j) \geq 0, \forall j \in \mc{N}_i,            \\
& \quad \vm{x}_i = \vm{z}_{ii}, \vm{x}_{j} = \vm{z}_{ij}, \forall	j \in \mc{N}_i, \quad \forall i \in \{1,...,N\}.
\end{aligned}
\end{equation}

The OA-ADMM augmented Lagrangian $\mc{L}_{\vm{\rho}}$ for this problem is then 
{\footnotesize \begin{equation} \label{eq:oacr_meth_MainLagrangian}
\begin{aligned}
\mathcal{L}_{\vm{\rho}}  = \ouset{N}{i=1}{\sum}\Big[  J_i(\vm{x}_i)+\vm{\lambda}_{ii}^\top (\vm{x}_i-\vm{z}_i)+ \Vert \vm{R}_{ii} (\vm{x}_i-\vm{z}_{ii}) \Vert^2_2  \\
+\  \underset{j \in \mathcal{N}_i}{\sum} \left( \vm{\lambda}_{ij}^\top ( \vm{x}_j - \vm{z}_{ij} ) + \Vert \vm{R}_{ij} (\vm{x}_j-\vm{z}_{ij})\Vert_2^2 \right) \Big],
\end{aligned}
\end{equation}}
where $\vm{\lambda}_{ii}$ (resp. $\vm{\lambda}_{ij}$) is the Lagrange multiplier in vector form for agent $i$ (resp. agent $i$ to agent $j$), and $\vm{R}_{ii}$ (resp. $\vm{R}_{ij}$) is the diagonal matrix form of $\vm{\rho}$ for agent $i$ (resp. agent $i$ to agent $j$).

We separate the overall optimization problem into smaller steps to enable the distribution of the computational load. The first step is the trajectory optimization (or $x$-update):
\begin{equation} \label{eq:oacr_meth_TrajUpdate}
\begin{aligned}
\vm{x}_i^{k+1}:= \arg \min_{\vm{x}_i\in \mc{X}_i^f} & \quad \mc{L}_{\vm{\rho},x,i}(\vm{x}_i,\vm{z}_{JI}^k,\vm{\lambda}^{k}_{JI},\vm{\rho}^k_{JI}), 
\end{aligned}
\end{equation}
where $\vm{z}_{JI}$ includes $\vm{z}_{ii}$ and $\vm{z}_{ji},\forall j\in \mc{N}_i$ , $\vm{\lambda}_{JI}$ is the combination of $\vm{\lambda}_{ii}$ and $\vm{\lambda}_{ji}, \forall j\in \mc{N}_i$, and $\vm{\rho}_{JI}$ includes $\vm{\rho}_{ii}$ and $\vm{\rho}_{ji}, \forall j \in \mc{N}_i$. Since the $x$-update only adjusts $\vm{x}_i$, we can use a lighter version of the full augmented Lagrangian in \eqref{eq:oacr_meth_MainLagrangian}:
{\footnotesize \begin{equation} \label{eq:oacr_meth_TrajLagrangian}
\begin{aligned}
\mc{L}_{\vm{\rho},x,i}	= J_i(\vm{x}_i)+{\vm{\lambda}^k_{ii}}^\top (\vm{x}_i-\vm{z}^k_{ii})+\Vert \vm{R}^k_{ii}(\vm{x}_i-\vm{z}^k_{ii})\Vert_2^2 \\
+ \underset{j \in \mathcal{N}_i}{\sum} \left( {\vm{\lambda}_{ji}^k}^\top ( \vm{x}_i - \vm{z}^k_{ji} ) +  \Vert \vm{R}^{k}_{ji}(\vm{x}_i-\vm{z}^k_{ji}) \Vert_2^2 \right),
\end{aligned}
\end{equation}}
where $\vm{\lambda}^k_{ji}$ (resp. $\vm{R}^{k}_{ji}$) is the Lagrange multiplier (resp. penalty matrix) for agent $j$ w.r.t agent $i$. The trajectory optimization step is fully parallelizable given that all the values in the augmented Lagrangian \eqref{eq:oacr_meth_TrajLagrangian} are either known or independent of other agents. Then, the resulting $\vm{x}_i^{k+1}$ has to be communicated to all nearby agents. After sending $\vm{x}_i^{k+1}$ to, and receiving $\vm{x}_j^{k+1}$ from, all $j\in\mc{N}_i$, the copy optimization step (or $z$-update) can be performed, which can be seen as the collision avoidance update because of the $d(\cdot) \geq 0$ constraint:
\begin{equation} \label{eq:oacr_meth_CopyUpdate}
\begin{aligned}
\vm{z}_{IJ}^{k+1}:= \arg \min_{\vm{z}_{IJ}} & \quad \mc{L}_{\vm{\rho},z,i}(\vm{x}_{I}^{k+1},\vm{z}_{IJ}^k,\vm{\lambda}^{k}_{IJ},\vm{\rho}^k_{IJ})                                  \\
s.t.                         & \quad d(\vm{z}_i^{k+1},\vm{z}_j^{k+1}) \geq 0,\ \forall j\in \mc{N}_i,
\end{aligned}
\end{equation}
where $\vm{z}_{IJ}$ (resp. $\vm{x}_{I}$) contains both $\vm{z}_{ii}$ and $\vm{z}_{ij},\forall j\in \mc{N}_i$ (resp. $\vm{x}_{i}$ and $\vm{x}_{j},\forall j\in \mc{N}_i$). The reduced augmented Lagrangian $\mc{L}_{\vm{\rho},z,i}$ for the $z$-update is defined as
{\footnotesize \begin{equation} \label{eq:oacr_meth_CopyLagrangian}
\begin{aligned}
\mc{L}_{\vm{\rho},z,i}= {\vm{\lambda}^k_{ii}}^\top (\vm{x}^{k+1}_i-\vm{z}_i)+ \Vert \vm{R}^k_{ii}( \vm{x}^{k+1}_i-\vm{z}_{ii}) \Vert_2^2 \\+ \underset{j \in \mathcal{N}_i}{\sum} \left( {\vm{\lambda}^k_{ij}}^\top ( \vm{x}^{k+1}_j - \vm{z}_{ij} ) + \Vert \vm{R}^{k}_{ij} ( \vm{x}^{k+1}_j - \vm{z}_{ij} ) \Vert_2^2 \right).
\end{aligned}
\end{equation}}

Following the $x$ and $z$ updates, the $\lambda$-update is performed:
\begin{equation} \label{eq:oacr_meth_LagrangianUpdate}
\begin{aligned}
\vm{\lambda}^{k+1}_{ii} := &\  \mu_{i}(\cdot) \vm{\lambda}^{k}_{ii}+\vm{\rho}_{ii}^{k}\circ(\vm{x}^{k+1}_i-\vm{z}_{ii}^{k+1}),                                   \\
\vm{\lambda}^{k+1}_{ij} := &\ \mu_{ij}(\cdot) \vm{\lambda}^{k}_{ij}+ \vm{\rho}_{ij}^{k}\circ(\vm{x}^{k+1}_j-\vm{z}_{ij}^{k+1}), \forall j\in \mc{N}_i	,
\end{aligned}
\end{equation}
where $\vm{\rho}_{ii}^{k+1},\vm{\rho}_{ij}^{k+1},\vm{z}_{ii}^{k+1}$, and $ \vm{z}_{ij}^{k+1}$ are all available locally, and $\mu$ is a forgetting factor representing the similarity between the system at the current and the previous time step. When multiple iterations of OA-ADMM are performed per control step, the value of $\mu$ can be assumed to be $1$ for all iterations in the same real-time time step. 

The final step of a single OA-ADMM iteration involves updating the penalty vector $\rho$:
\begin{equation} \label{eq:oacr_meth_AdaptationUpdate}
\begin{aligned}
\vm{\rho}^{k+1}_{ij} := &\ \phi_{ij}(\vm{x}_i^{k+1},\vm{x}_j^{k+1}),\forall j\in \mc{N}_i \\
\vm{\rho}^{k+1}_{ii} := &\ \phi_{ii}(\vm{x}_i^{k+1}),\\
\end{aligned}
\end{equation}
where $\phi(\cdot)$ is a function of the states, whose expression can be chosen depending on the desired behavior, provided that the resulting $\vm{R}$ is always a positive definite diagonal matrix. The updated values of $\vm{z}^{k+1}$, $\vm{\rho}^{k+1}$, and $\vm{\lambda}^{k+1}$ are then communicated to complete one OA-ADMM iteration.

\begin{figure}[t]
	\centering
	\resizebox{0.9\linewidth}{!}{
		\begin{tikzpicture}[auto, node distance=4.5cm] 
		\node [block, name=x-update,align=center,minimum height = 1.35cm,minimum width = 2.2cm] {$x$-update:\\ \Cref{eq:oacr_meth_TrajUpdate}};
		\node [block, right= 0.8cm of x-update, name=z-update, align=center,minimum height = 1.35cm,minimum width = 2.2cm] {$z$-update: \\ \Cref{eq:oacr_meth_CopyUpdate}};
		\node [block, right= 1.2cm of z-update, name=lambda-update,align=center,minimum height = 1.35cm,minimum width = 2.2cm] {$\lambda$-update: \\ \Cref{eq:oacr_meth_LagrangianUpdate}};
		\node [block, right=0.8cm of lambda-update, name= rho-update, align=center,minimum height = 1.35cm,minimum width = 2.2cm] {$\rho$-update:\\ \Cref{eq:oacr_meth_AdaptationUpdate}};
		\node [block, below= 0.75cm of x-update, name=commX,align=center,minimum height = 1.0cm,minimum width = 2.2cm] {communicate} ;
		\node [block, above= 1.85cm of x-update, name=comm2,align=center,minimum height = 1.0cm,minimum width = 2.2cm] {communicate} ;
		\node [mux, right=1.35cm of comm2, xshift= -0.5cm, name=mux] {};
		\node [block, above= 0cm of x-update, name=followX,align=center,minimum height = 0.0cm,minimum width = 2.2cm] {MPC($\vm{x}^{k}$)};
		\node [block, below= 0cm of comm2, name=kUpdate,align=center,minimum height = 0.0cm,minimum width = 2.2cm] {$k = k+1$};	
		
		\draw [->] (x-update) -- node[] {$\vm{x}_i^{+}$} (z-update);
		\draw [->] (x-update) -- node[yshift = 0.1cm] {$\vm{x}_i^{+}$} (commX);
		\draw [->] (commX) -| node[yshift = -0.35cm,xshift=0.15cm] {$\vm{x}_j^{+},\forall j\in\mc{N}_i$} (z-update);
		
		\draw [->] (z-update) -- node[] {$\vm{z}_{IJ}^{+}$} (lambda-update);
		\draw [->] ([xshift=0.25cm]z-update.north) -- node[right] {$\vm{z}^{+}_{IJ}$} ++(0,1cm) |- ([yshift=-0.2cm] mux.east);
		\draw [->] ([xshift=-0.35cm]lambda-update.north) -- node[left] {$\vm{\lambda}^{+}_{IJ}$} ++(0,1cm) |- ([yshift=0.2cm] mux.east);
		\draw [->] (rho-update.north) -- node[right] {$\vm{\rho}^{+}_{IJ}$} ++(0,1cm) -| ([xshift=0.35cm] lambda-update.north);
		
		\draw [->] (mux.west) -- (comm2.east);	
		\draw [dotted,->] (lambda-update) -- (rho-update); 
		
		\draw [->] (kUpdate.south) -- node[right,yshift = -0.25cm] {$\vm{z}^k_{JI},\vm{\lambda}^{k}_{JI}$} ++(0,0) -- (followX);	
		\end{tikzpicture}%
	}
	\caption{OA-ADMM based conflict resolution algorithm for agent $i$. We use the notation $\vm{x}^{+}:=\vm{x}^{k+1}$. The dotted arrow indicates code execution order; variables are automatically perpetuated along the arrows, e.g. the $\lambda$-update block receives $\vm{x}^+_j$ via the $z$-update block.}%
	\label{fig:oa-admm_meth_Diagram2}%
	\vspace{-20pt}
\end{figure}
  \subsection{Designing the adaptation function $\phi(\cdot)$} \label{ssec:oa-admm_meth_DesignPhi}	
	Possible approaches to design adaptive penalty parameters for faster convergence have been explored in \cite{he_alternating_2000} and \cite{xu_adaptive_2017}. With OA-ADMM, we may want to fit other needs. For example, in the case of real-time optimization for decentralized conflict resolution, we prioritize safety over convergence speed. Furthermore, instead of using rule-based adaptation schemes as in \cite{he_alternating_2000} and \cite{xu_adaptive_2017}, we use a more general adaptation function $\phi$.
	
	The requirements on $\phi(\cdot)$ for online convergence can be summarized as $\dot{\phi}(\cdot) = 0, \txtn{for }t\rightarrow\infty$. In addition of the basic requirement on $\phi$, we also have to take into account the purpose of $\phi$ in the real-time control case.  Since we are applying OA-ADMM and MPC to a motion planning problem involving autonomous vehicles, we wish to use $\phi$ to improve the online collision avoidance behavior. The value of the penalty parameter $\vm{\rho}$ has a large influence on the convergence rate of OA-ADMM; additionally, $\vm{\rho}$ tunes the importance of the primal and dual residuals during optimization, with a large $\vm{\rho}$ prioritizing the primal residual $\vm{r}^k$, and a small $\vm{\rho}$ prioritizing the dual residual $\vm{s}^k$. 
	
	Given that the primal residual for \eqref{eq:oacr_meth_slack} is $\vm{x}-\vm{z}$, increasing $\vm{\rho}$ effectively increases the penalty for the actual trajectory deviating from the copies. However, since \eqref{eq:oacr_meth_slack} is a multi-agent problem, there are also agent specific values of $\vm{\rho}$. In essence, each agent $i$ has three relevant types of $\vm{\rho}$, namely $\vm{\rho}_{ii}$, $\vm{\rho}_{ij}$ and $\vm{\rho}_{ji}$. The first type, $\vm{\rho}_{ii}$, directly affects $\vm{\lambda}_{ii}^{k+1}$ by scaling part of the primal residual ($\vm{x}^{k+1}_{ii} - \vm{z}^{k+1}_{ii}$), it is also present in the augmented Lagrangian, acting as a weight on the residual inside the squared L2 norm. The value of $\vm{\rho}_{ii}$ can therefore be summarized as the weight for the cost of the deviation between $\vm{x}^{k+1}_{ii}$ (resp. $\vm{z}^{k+1}_{ii}$) and $\vm{z}^k_{ii}$ (resp. $\vm{x}^{k+1}_{ii}$). The second type of $\vm{\rho}$ is $\vm{\rho}_{ij}$, which is present in the calculation of $\vm{\lambda}^{k+1}_{ij}$ as a weight for the residual $\vm{x}^{k+1}_k-\vm{z}^{k+1}_{ij}$ and in the augmented Lagrangian $\mc{L}_{\vm{\rho},z,i}$ in which it scales the same residual inside the norm. As a result, the value of $\vm{\rho}_{ij}$ directly affects the $z$-update for agent $i$, with larger values allowing less deviation of $\vm{z}^{k+1}_{ij}$ from $\vm{x}^{k+1}_j$. The final $\vm{\rho}$ is $\vm{\rho}_{ji}$, which is in essence the reverse of $\vm{\rho}_{ij}$, i.e. $\vm{\rho}_{1,2}=\vm{\rho}_{2,1}$. From the perspective of agent $i$, $\vm{\rho}_{ji}$ acts purely in the $x$-update, penalizing deviation of $\vm{x}^{k+1}_i$ from $\vm{z}^{k+1}_{ji}$. 
	
	The adaptation function used in this paper, to enhance online safety and robustness, adapts the penalty parameter based on the physical states:
    \begin{equation}\footnotesize \label{eq:sim_imp_oa-admm_Phi_ij}
    	\phi(\cdot)_{ij} = \begin{cases} w_i \vm{\phi}^{min} , \txtn{ if } \left( \frac{D}{d(\vm{x}_i^k,\vm{x}_j^k)}\right)^a < \vm{\phi}^{min} \\
    		w_i \vm{\phi}^{max}, \txtn{ if } \left( \frac{D}{d(\vm{x}_i^k,\vm{x}_j^k)}\right)^a > \vm{\phi}^{max} \\
    		w_i \left( \frac{D}{d(\vm{x}_i^k,\vm{x}_j^k)}\right)^a, \txtn{ otherwise }\\
    	\end{cases}
    \end{equation}	
    and 
    \begin{equation} \label{eq:sim_imp_oa-admm_Phi_ii}
    	\phi(\cdot)_{ii} = w_i \frac{1}{N_i} \sum_{j \in \mc{N}_i}\left( \phi(\cdot)_{ij}\right)^a,
    \end{equation}	
	where $D$ is the minimum distance, $w_i$ is a weight that can modify the importance of agent $i$, $a$ is a variable that determines the shape of $\phi$, and $N_i$ is the amount of agents in $\mc{N}_i$. The minimum bound $\vm{\phi}^{min}$ ensures that $\rho>0$, avoiding the problem becoming ill-conditioned, the maximum bound $\vm{\phi}^{max}$ limits the possibility for extreme values of $\phi$, which can destabilize the system. For \eqref{eq:sim_imp_oa-admm_Phi_ij}, this can occur when $d(\vm{x}_i,\vm{x}_j)\approx 0$, which can occur when planned trajectories overlap. 
    The $\phi$ given in \eqref{eq:sim_imp_oa-admm_Phi_ij}-\eqref{eq:sim_imp_oa-admm_Phi_ii} therefore adjusts the value of $\rho_{ij}$ and $\rho_{ii}$ when the online MPC is planning trajectories with high probability of collisions, increasing the primal feasibility whilst sacrificing individual optimality. The advantage of this method is that this distance-based approach is simple to design, yet achieves results similar to more advanced techniques. The function given in \eqref{eq:sim_imp_oa-admm_Phi_ij}-\eqref{eq:sim_imp_oa-admm_Phi_ii} can be interpreted as a control barrier or potential field function for the multi-agent motion planning problem.
  \subsection{Designing the similarity function $\mu(\cdot)$ } \label{ssec:croa_meth_DesignMu}
  	When $\mu(\cdot) = 1$, the change from $t$ to $t+\delta t$ has no effect on the previous OA-ADMM iterations, i.e. $\vm{\lambda}^{k+1} = \vm{\lambda}^{k} + \vm{\rho}^k \circ \vm{r}^{k+1}$. Conversely a $\mu(\cdot)=0$ implies that there is no useful relation between the previous time step $t$ and the current time step $t+\delta t$, i.e. $\vm{\lambda}^{k+1} = \vm{\rho}^k \circ \vm{r}^{k+1}$. The difficulty is however designing a function $\mu$ that, using the information available, results in effective online performance. If the system is fully known, it might be possible to analytically find the optimal $\mu^\star$, this is however a time intensive procedure and very system dependent. Instead we attempt an intuitive approach to find a $\mu$ which approximates the behavior of $\mu^\star$.
  	
  	We know that if $\mu(\cdot)^\star =1$ it should hold that $\vm{x}^\star(t) = \vm{x}^\star(t+\delta t)$: the optimum should not change from $t$ to $t+\delta t$ if $\mu=1$. Additionally, if OA-ADMM has reached the optimum, then $\vm{x}^k(t) = \vm{x}^{k+1}(t)$. Ergo, if OA-ADMM has converged and $\mu^\star =1$, then $\vm{x}^k(t) = \vm{x}^{k+1}(t+\delta t).$	Given that this can be calculated in run-time, we can utilize this to construct a $\mu$ which approximates $\mu^\star$ in the optimum. For example, the following formula satisfies the requirements:
	\begin{equation} \label{eq:croa_meth_dsgn_muApprox1}
		\mu(\cdot) = w_x \left( 1-\frac{\Vert \vm{x}^{k+1}(t+\delta t) - \vm{x}^{k}(t) \Vert_2 }{\Vert \vm{x}^{k}(t) \Vert_2 }\right),
	\end{equation}
	where $w_x + w_z + w_{\lambda} + w_{\rho} = 1$. Note that this also hold for $\vm{z},\vm{\lambda},$ and $\vm{\rho}$. This approach, however, requires that OA-ADMM is performed until convergence, as only then $\vm{x}^\star(t) = \vm{x}^\star(t+\delta t)$ is guaranteed.
		
	Another approach is to construct a $\mu(\cdot)$ by evaluating the role of $\mu$. The similarity function implemented is based on the idea that the relevance of the previous $\vm{\lambda}$ is positively correlated to the value of $\vm{\rho}$. An intuitive explanation for the conflict resolution case is to view $\vm{\lambda}$ as a penalty by OA-ADMM aiming to enforce the collision avoidance constraint. When a collision is likely, $\vm{\rho}$ will increase due to the design of the $\phi$. In this case, it is desirable to increase the penalty $\vm{\lambda}$ to enforce the constraint. However, when collision are unlikely, continuing with the previous $\vm{\lambda}$ can result in suboptimality. This concept is implemented as follows:
    \begin{equation}
       \mu(\cdot,k) = \eta \mu(\cdot,k-1) + (1-\eta) \txtn{min}(\rho_{JI}\frac{1}{w_i},\vm{1}),
    \end{equation}
    where the elements of $\mu$ are bounded to be less or equal to one, along with a weighted average (scaled with $0\leq\eta\leq1$ acting as a simple filter to reduce the effects of disturbances) between the current and the previous value of $\mu$.

\section{Numerical Simulations}
In this section, we evaluate the robustness of OA-ADMM MPC for an autonomous vehicle simulated in CARLA, additionally we compare the conflict resolution efficiency against the decentralized conflict resolution methods from \cite{azimi_reliable_2013} (AMP-IP) and \cite{liu_distributed_2018} (TDCR\footnote{The conflict resolution method proposed in \cite{liu_distributed_2018} is unnamed, for convenience sake we will refer to it as the Timeslot-based Decentralized Conflict Resolution method (TDCR).}). Both methods are limited in terms of control input, both only able to adjust their velocities along the planned trajectory. Whilst AMP-IP is a reactive strategy, TDCR uses prediction in their method, which allows vehicles to plan their velocities ahead accordingly. 

\subsection{Simulation Setup}
\begin{figure}[ht]
    \vspace{-10pt}
	\centering
	\includegraphics[width=0.725\linewidth]{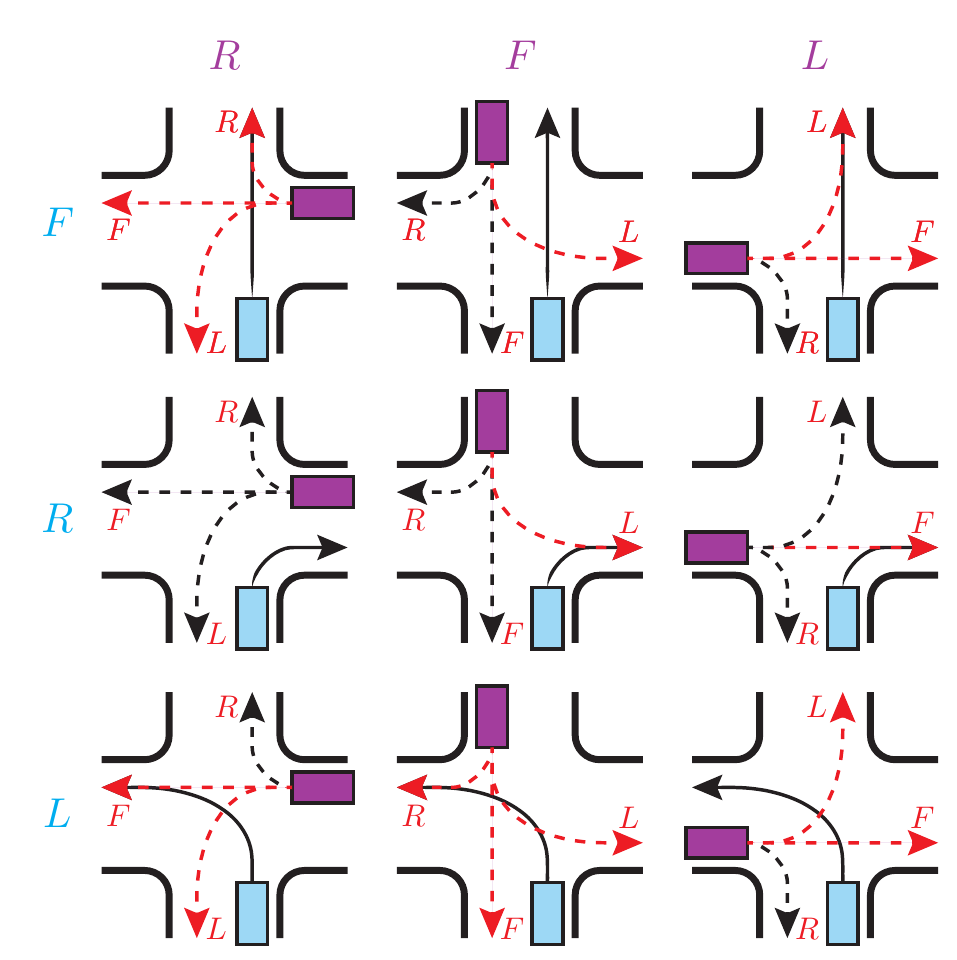}
	\caption{Conflict cases used for the benchmark. Solid lines indicate the desired path for the blue vehicle and are marked with blue letters, the relative direction of the magenta vehicle w.r.t. the blue vehicle is marked with magenta letters, with their relative actions marked with red letters. }
	\label{fig:si_benchmarks_MOMD_PossibleConflicts}
	\vspace{-10pt}
\end{figure}

The simulations are carried out using the benchmarking tool described in \Cref{apx:Benchmark}.
The metrics measured are the total travel time per vehicle for their respective cases, these are compared against the no conflict case for each respective protocol to get the added delay caused by each protocol. The no conflict case for each protocol simulates the same amount of vehicles with the same exact reference velocities, ensured by the identical random seeds. A major difference between OA-ADMM MPC and the traditional methods of AMP-IP and TDCR lies in that OA-ADMM MPC does not require the map of the environment beforehand. To attempt to show the effects of this prior knowledge, the traditional approaches are simulated for the three different fidelity cases: a 1x1 grid (low fidelity), a 4x4 grid (medium fidelity), and a 8x8 grid (high fidelity), with all grids having a dimensions of 18x18m centered at the intersection.
All the protocols are tested by spawning vehicles at equal distance to the intersection center with a reference velocity of $4$~m/s, with uniformly distributed variations between $-0.15$~m/s and $0.15$~m/s. All the possible cases depicted in \Cref{fig:si_benchmarks_MOMD_PossibleConflicts}, except for the vehicle follower case, are simulated in threefold and averaged out. The simulator is ran at a frequency of 160~Hz, with the vehicles running the protocols at 20~Hz; vehicles perform the low level control at 40~Hz to reduce instability. 
\subsection{Simulation Results}
\begin{figure}[ht]
	\centering
	\resizebox{\linewidth}{!}{
%
%
\begin{tikzpicture}
\begin{axis}[%
	width=1.65\linewidth,
	height=1.65*0.35\linewidth,
	scale only axis,
	xmin=0.50,
	xmax=7.50,
	xtick={1,2,3,4,5,6,7},
	xticklabels={{AMP-IP (l)},{AMP-IP (m)},{AMP-IP (h)},{TDCR (l)},{TDCR (m)},{TDCR (h)},{OA-ADMM}},
	ymin=0,
	ymax=4.5,
	ylabel={Mean Delay (s)},
	axis background/.style={fill=white},
	axis x line*=bottom,
	axis y line*=left,
	ymajorgrids=true,
    grid style=dashed
	]
	\addplot[ybar, bar width=0.55cm, draw=black, fill=clrAMPIP] coordinates { (1, 4.002) };
	\addplot[ybar, bar width=0.55cm, draw=black, fill=clrAMPIP!90] coordinates {(2, 1.542) };
	\addplot[ybar, bar width=0.55cm, draw=black, fill=clrAMPIP!80] coordinates { (3, 1.427) };
	\addplot[ybar, bar width=0.55cm, draw=black, fill=clrTDCR] coordinates { (4, 3.684) };
	\addplot[ybar, bar width=0.55cm, draw=black, fill=clrTDCR!90] coordinates { (5, 1.307) };
	\addplot[ybar, bar width=0.55cm, draw=black, fill=clrTDCR!80] coordinates { (6, 1.165) };
	\addplot[ybar, bar width=0.55cm, draw=black, fill=clrOAADMM] coordinates { (7, 0.802) };
\end{axis}
\end{tikzpicture}%
	}
	\caption{Mean delay values for the decentralized protocols.}
	\label{fig:sim_res_Bar}
	\vspace{-10pt}
\end{figure}
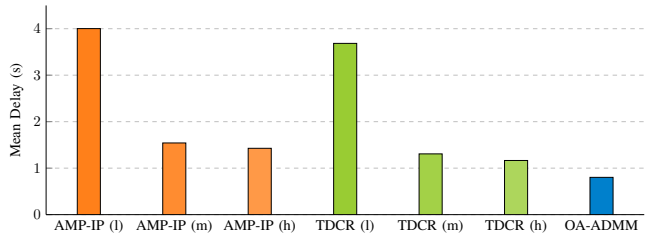
The average delay for the AMP-IP, TDCR, and OA-ADMM is shown in \Cref{fig:sim_res_Bar}, where AMP-IP and TDCR are shown separately for each of their grid fidelity cases. 

\begin{table}[ht]
	\centering	
	\resizebox{0.85\linewidth}{!}{
	\begin{tabular}{@{}lp{12mm}p{12mm}p{15mm}@{}}
		\toprule
		Protocol (case)               & Mean \newline Time (s) & Mean \newline Delay (s) & Mean Add. \newline Delay (s) \\ \midrule
		AMP-IP (n)         & 17.804                 & ($-$)             & ($-$)               \\
		AMP-IP (l)         & 21.806                 & 4.002             & 3.593               \\
		AMP-IP (m)\quad    & 19.346                 & 1.542             & 1.133               \\
		AMP-IP (h)         & 19.231                 & 1.427             & 1.018               \\ \midrule
		TDCR (n)           & 17.809                 & ($-$)             & ($-$)               \\
		TDCR (l)           & 21.493                 & 3.684             & 3.275               \\
		TDCR (m)           & 19.117                 & 1.307             & 0.898               \\
		TDCR (h)           & 18.974                 & 1.165             & 0.756               \\ \midrule
		OA-ADMM (n)        & 18.158                 & ($-$)             & ($-$)               \\
		OA-ADMM                       & 18.960                 & 0.802             & 0.394               \\ \bottomrule
	\end{tabular}}
	\caption{Summary of AMP-IP, TDCR, and OA-ADMM results. The mean delay is measured relative to the no conflict case, the mean added delay is measured against the mean estimated delay (0.4089~s) from the benchmark.} 	\label{tbl:sim_res_Times}
	\vspace{-5pt}
\end{table}

The mean time and mean delay for all the protocols are given in \Cref{tbl:sim_res_Times}. Compared with AMP-IP, OA-ADMM MPC is found to have a 79.95\%, 47.95\%, and 43.74\% decrease in mean delay for the low, medium, and high fidelity cases respectively. The percentage decrease in average added delay regarding the low, medium, and high fidelity cases for AMP-IP are 89.04\%, 65.25\%, and 61.32\% respectively.
Compared with TDCR, OA-ADMM MPC is found to have a 78.21\%, 38.61\%, and 31.11\% decrease in mean delay for the low, medium, and high fidelity cases respectively.  The percentage decrease in mean added delay regarding the low, medium, and high fidelity cases for TDCR are 87.98\%, 56.18\%, and 47.93\% respectively. 

A detailed overview of the delays for the protocols is given in \Cref{apx:sim_res_DetailedTimes}, where the delay for each individual conflict case is given separately. Note that only the high fidelity cases are shown, the delays for the lower fidelity cases are deemed less relevant for the detailed comparison as they are always higher than the delays for the high fidelity cases.  

The results indicate that OA-ADMM MPC is outperforming both conventional methods for all cases. This can be attributed to the use of a collision avoidance constraint compared with an entry time constraint used by TDCR. An entry time constraint is limited in detail by the size of the cells, thereby effective conflict resolution requires detailed prior knowledge of the environment, which can be costly to obtain/save and is not always available. OA-ADMM MPC, however, only requires the relative positions of the vehicles, which can be obtained in real-time with relative ease.

\subsection{Tuning Complexity}
In addition to the simulations conducted in CARLA using the benchmark, the tuning complexity of OA-ADMM compared with ADMM is analyzed using a simpler MATLAB test case. The test case involves four holonomic circular robots approaching an intersection simultaneously. To avoid deadlocks, vehicles on the horizontal lane have their values of $w_i$ from \eqref{eq:sim_imp_oa-admm_Phi_ij} doubled.
\begin{table}[ht]
	\centering	
	\resizebox{0.9\linewidth}{!}{
	\begin{tabular}{@{}lp{3mm}p{6mm}p{5mm}p{11mm}p{15mm}@{}}
		\toprule
		Algorithm      & Time\newline outs & Viol-\newline ations & Re-\newline solved & Mean \newline Delay (s) & Mean \newline MSV (m$^2$) \\ \midrule
		ADMM         & 86 & 133 & 1 & 5.35 & 2.33$\cdot10^{-2}$\\
		OA-ADMM        & 49 & 83 & 88 & 2.93 & 8.83$\cdot10^{-3}$\\
 \bottomrule
	\end{tabular}}
	\caption{Comparison between ADMM and OA-ADMM for 220 hyperparameter combinations. Timeouts
imply that the case did not resolve within 30 seconds; violations are cases with
constraint violations; resolved implies that no constraint violation or timeout has
occurred.}	\label{tbl:sim_res_Tuning}
	\vspace{-5pt}
\end{table}

To gain insight on the tuning complexity of OA-ADMM, the simulation is performed for a range of hyperparameter combinations: $D\in \{0,0.1,...,1\}$ and $w_i\in \{0.25,0.5,...,5\}$. These combinations are then simulated and evaluated for delay and mean square constraint violations (MSV), the results of which are given in \Cref{tbl:sim_res_Tuning}. The results indicate that OA-ADMM, for the tested cases, resolves significantly more cases, whilst having shorter delays. In the cases where the constraint was violated, OA-ADMM had lower values of MSV, indicating that OA-ADMM is significantly easier to tune and more robust then conventional ADMM when used in combination with MPC. 


\section{Conclusion} \label{ssec:dis_Conclusion}
OA-ADMM is a novel, flexible framework to use ADMM for robust online optimization. In our case study, the chosen adaptation function improves the robustness of decentralized MPC enough to achieve improved conflict resolution efficiency compared with competing decentralized conflict resolution methods like AMP-IP and TDCR.

Given that OA-ADMM is a novel framework, a lot of work can be done to further explore the proposed adaptation function and similarity function, including e.g. the possibility of an optimal similarity function. Better guarantees for online convergence of OA-ADMM could be provided in combination with stricter requirements on the online system and the adaptation and similarity functions.


\bibliography{MyBib,ManualBib}

\newpage

\appendices
\crefalias{section}{appendix}

\section{Proof of \Cref{th:oa-admm_cprf_stat_Theorem}} \label{apx:proof_theorem1}
\begin{proof}
Iterating \Cref{lm:oa-admm_cprf_static_Lem1} from $k=0$ to $\infty$ gives
$\ouset{\infty}{k=0}{\sum} \left( \Vert \vm{R}^k \vm{r}^{k+1}\Vert_2^2 + \Vert \vm{R}^k \vm{B}(\vm{z}^{k+1}-\vm{z}^k)\Vert_2^2  \right) \leq V^0,$
which simply states that for $k=\{0,...,\infty\}$ the sum of the Lyapunov function is bounded, implying that $V^k \rightarrow 0$ as $k\rightarrow \infty$. Given that $V$ is a sum of two squared L2-norms, it has to hold that both $\vm{R}^k \vm{r}^{k+1} \rightarrow 0$ and $\vm{R}^k \vm{B}(\vm{z}^{k+1}-\vm{z}^k)\rightarrow 0$. Because $\vm{R}$ is a symmetric positive definite matrix, it also holds that $\vm{r}^{k+1} \rightarrow 0$ and $\vm{B}(\vm{z}^{k+1}-\vm{z}^k)\rightarrow 0$, i.e. the primal and dual residuals converge to zero. \Cref{lm:oa-admm_cprf_static_Lem} provides bounds for the objective suboptimality $p^{k+1}-p^\star$ ensuring that it converges to zero as the residuals converge to zero, hence as $k\rightarrow \infty$. \Cref{lm:oa-admm_cprf_static_Lem4} allows the use of static $\vm{\rho}$ convergence results as long as the dynamic $\vm{\rho}$ converges.
\end{proof}

\section{Added Delay Conflict Resolution Benchmark} \label{apx:Benchmark}
To compare protocols against each other in CARLA, a common benchmark has to be used.\footnote{Available at {\tt\small https://github.com/jerryangit/\\AddedDelayCRBenchmark}} In order for the benchmark to provide a reference point for the results, an estimated delay is desired. The theoretical estimated delay for a certain case however is difficult to directly calculate mainly due to the nonlinear nature of the agent dynamics, the large combinatorial passing order problem, etc. Therefore, some assumptions have to be made to calculate a usable metric; we will investigate the delay when there are two vehicles arriving at the intersection at once. This allows a direct comparison between the travel time for certain conflict resolution protocols in an easy to interpret manner. The main metrics will therefore be travel time and delay, where the delay is measures compared with the no conflict scenario: only one vehicle traverses the intersection at once, corresponding to the minimum travel time. 

Following the zero delay assumption, the only segment of a trajectory contributing to the delay is the one where there is an orthogonal component to the trajectory with respect to the trajectory of the yielding vehicle. The delay caused can then be calculated by taking the displacement function along this axis and dividing it by the velocity function along this axis. By comparing measured delay against the estimated delay, the added delay metric can be found.

\section{Additional Results} \label{apx:sim_res_DetailedTimes}
\begin{table}[ht]
	\centering
	\resizebox{\linewidth}{!}{
    	\begin{tabular}{@{}llllllllll@{}}
	\toprule
	             & L,L & L,F  & L,R  & F,L  & F,F  & F,R  & R,L  & R,F  & R,R  \\ \midrule
	Est. Delay L & 0.80 & 0.80 & 0.00 & 0.00 & 0.72 & 0.83 & 0.80 & 0.75 & 0.00 \\
	Est. Delay F & 0.75 & 0.66 & 0.00 & 0.72 & 0.00 & 0.00 & 0.80 & 0.66 & 0.96 \\
	Est. Delay R & 0.00 & 0.96 & 0.00 & 0.83 & 0.00 & 0.00 & 0.00 & 0.00 & 0.00 \\ \midrule
	AMP-IP (h) L & 1.66 & 1.55 & 0.86 & 1.15 & 1.40 & 1.33 & 1.66 & 1.92 & 0.00 \\
	AMP-IP (h) F & 1.92 & 0.59 & 0.00 & 1.40 & 0.00 & 0.00 & 1.55 & 0.59 & 1.01 \\
	AMP-IP (h) R & 0.00 & 1.01 & 0.00 & 1.33 & 0.00 & 0.00 & 0.86 & 0.00 & 0.00 \\ \midrule
	TDCR (h) L   & 1.26 & 1.60 & 1.22 & 1.31 & 1.45 & 0.35 & 1.26 & 1.21 & 0.00 \\
	TDCR (h) F   & 1.21 & 0.56 & 0.00 & 1.45 & 0.00 & 0.00 & 1.60 & 0.56 & 0.20 \\
	TDCR (h) R   & 0.00 & 0.20 & 0.00 & 0.35 & 0.00 & 0.00 & 1.22 & 0.00 & 0.00 \\ \midrule
	OA-ADMM L    & 0.71 & 0.55 & 0.00 & 0.02 & 0.15 & 0.00 & 0.71 & 1.91 & 0.00 \\
	OA-ADMM F    & 1.91 & 0.69 & 0.00 & 0.15 & 0.00 & 0.00 & 0.55 & 0.69 & 0.02 \\
	OA-ADMM R    & 0.00 & 0.02 & 0.01 & 0.00 & 0.01 & 0.00 & 0.00 & 0.00 & 0.01 \\ \bottomrule
\end{tabular}
	}	
	\caption{Detailed table of AMP-IP, TDCR, and OA-ADMM results for the Carla simulations, showing the delays for the cases given in \Cref{fig:si_benchmarks_MOMD_PossibleConflicts}.} 	\label{tbl:sim_res_DetailedTimes}
	\vspace{-19pt}

\end{table}
\end{document}